\documentclass[conference]{IEEEtran}
\IEEEoverridecommandlockouts
\usepackage{subfig}
\usepackage{cite}
\usepackage{amsmath,amssymb,amsfonts}
\usepackage{algorithmic}
\usepackage{graphicx}
\usepackage{textcomp}
\usepackage{xcolor}
\usepackage{hyperref}
\usepackage{subfig}
\usepackage{threeparttable} 

\def\BibTeX{{\rm B\kern-.05em{\sc i\kern-.025em b}\kern-.08em
    T\kern-.1667em\lower.7ex\hbox{E}\kern-.125emX}}

\usepackage{tikz}
\newcommand*\circled[1]{\tikz[baseline=(char.base)]{
            \node[shape=circle,draw,inner sep=0.5pt] (char) {#1};}}

\usepackage{listings}

\definecolor{codegreen}{rgb}{0,0.6,0}
\definecolor{codegray}{rgb}{0.35,0.35,0.35}
\definecolor{codepurple}{rgb}{0.58,0,0.82}
\definecolor{backcolour}{rgb}{0.95,0.95,0.92}
\lstdefinestyle{mystyle}{
  backgroundcolor=\color{backcolour}, commentstyle=\color{codegreen},
  keywordstyle=\color{magenta},
  numberstyle=\tiny\color{codegray},
  stringstyle=\color{codepurple},
  basicstyle=\ttfamily\footnotesize,
  breakatwhitespace=false,         
  breaklines=true,                 
  captionpos=b,                    
  keepspaces=true,                 
  numbers=left,                    
  numbersep=5pt,                  
  showspaces=false,                
  showstringspaces=false,
  showtabs=false,                  
  tabsize=2
}

\begin{document}

\title{Accelerating Edge AI with Morpher: An Integrated Design, Compilation and Simulation Framework for CGRAs \thanks{This work was accepted by the Workshop on Compilers, Deployment, and Tooling for Edge AI (CODAI 2023), co-hosted at Embedded Systems Week on September 21st, 2023.}
}

\author{\IEEEauthorblockN{Dhananjaya Wijerathne, Zhaoying Li, Tulika Mitra}
\IEEEauthorblockA{\textit{School of Computing, National University of Singapore} \\
\{dmd, zhaoying, 
tulika\}@comp.nus.edu.sg }}

\maketitle

\begin{abstract}
Coarse-Grained Reconfigurable Arrays (CGRAs) hold great promise as power-efficient edge accelerator, offering versatility beyond AI applications. Morpher, an open-source, architecture-adaptive CGRA design framework, is specifically designed to explore the vast design space of CGRAs. The comprehensive ecosystem of Morpher includes a tailored compiler, simulator, accelerator synthesis, and validation framework. This study provides an overview of Morpher, highlighting its capabilities in automatically compiling AI application kernels onto user-defined CGRA architectures and verifying their functionality. Through the Morpher framework, the versatility of CGRAs is harnessed to facilitate efficient compilation and verification of edge AI applications, covering important kernels representative of a wide range of embedded AI workloads.
Morpher is available online at \url{https://github.com/ecolab-nus/morpher-v2}.

\end{abstract}


\section{Introduction}
In the ever-evolving era of artificial intelligence, the growing need for edge devices to adeptly manage advanced machine learning (ML) workloads is becoming increasingly important. 
These devices, operating under severe power and computational performance constraints, must not only efficiently execute a wide range of ML algorithms, but also cater to an array of diverse workloads such as signal and image processing. 
Even within the ML sphere, the advent of new kernel types presents a continuous challenge, outpacing the capabilities of current ML accelerators. 
Despite their efficiency in traditional ML workloads, these accelerators fall short in adaptability for non-ML tasks, highlighting the urgent need for more flexible solutions.


Emerging as a solution to this, Coarse-Grained Reconfigurable Arrays (CGRAs) represent an innovative class of hardware accelerators, combining the flexibility of FPGAs with a higher energy efficiency comparable to ASIC-based ML accelerators.
Their inherent word-level reconfigurability and high energy efficiency make them ideal for power and area-constrained edge devices. Additionally, the use of dataflow computing in CGRAs naturally aligns with the computational patterns of many AI workloads. 
Their unique blend of adaptability and efficiency has led to adoption of CGRAs commercially in Samsung Exynos 7420 SoC~\cite{samsungexy}, Intel Configurable Spatial Accelerator~\cite{intelspatial}, Sambanova  RDU~\cite{plasticine}, Renesas Configurable Processor~\cite{renesas}, and academic prototypes like HyCUBE\cite{karunaratne2017hycube, wangbo2019hycube} among others~\cite{dyser, adres, farahini, cascade}. 


\begin{figure}[t]
	\scriptsize
	\centering
	\includegraphics[width=0.6\linewidth]{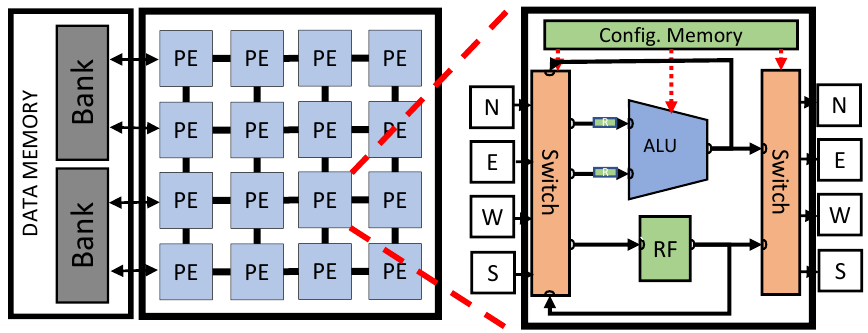}
	\caption{A 4x4 CGRA Architecture}
	\label{fig:cgra}
	\vspace{-.5cm}
\end{figure}

A CGRA architecture, as depicted in Figure~\ref{fig:cgra}, is characterized by a grid of interconnected Processing Elements (PEs) and multi-banked memories accessible to a subset of PEs, rendering it simple yet robust. 
The PEs consist of configurable switches, a register file, ALU, and control memory, enabling the time-multiplexed execution of instructions. 
The use of static scheduling negates the need for hardware structures for conflict resolution and synchronization, leading to a lightweight footprint for CGRAs. However, the effectiveness of CGRAs relies heavily on high-quality compiler mapping of application kernels onto the architecture, marking the CGRA compilation problem a substantial research area~\cite{li2022lisa, cgrame, mei2002dresc, wijerathne2021himap, wijerathne2022panorama, pillars, opencgra, ccf}.

\begin{figure}[t]
	\centering
	\subfloat[]{\includegraphics[width=.15\linewidth]{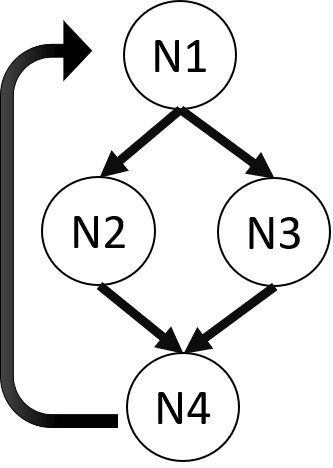}\label{fig:bg_dfg}}
	\subfloat[]{\includegraphics[width=.26\linewidth]{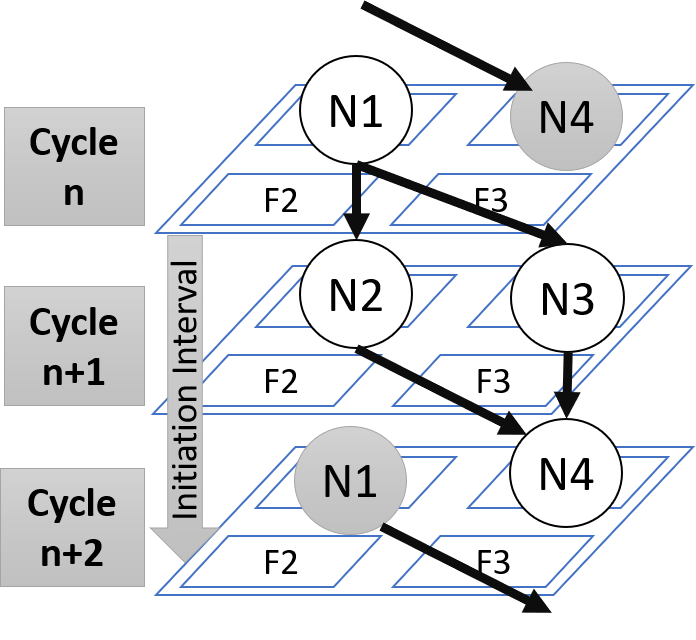}\label{fig:bg_sch}}
	\subfloat[]{\includegraphics[width=.4\linewidth]{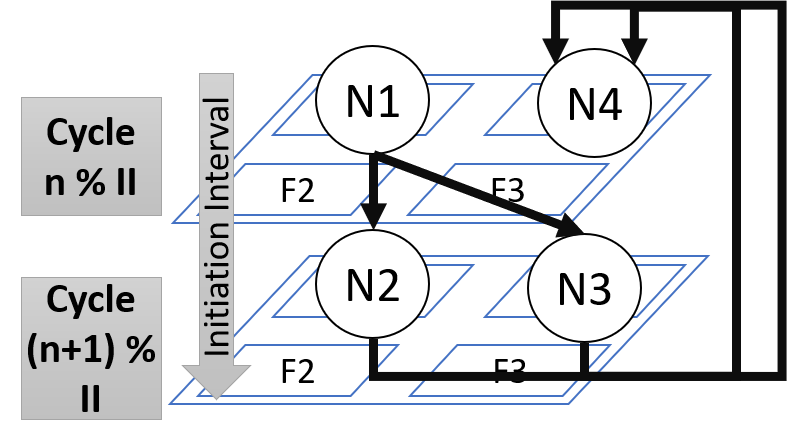}\label{fig:bg_mrrg}}
	\caption{ (a) a loop DFG (b) a loop schedule (c) a loop scheduled on MRRG (Modulo Resource Routing Graph).}
	\label{fig:bg_map}
	\vspace{-.3cm}
\end{figure}

Application kernels are typically statically scheduled on CGRAs, a process that exposes all architectural features to the compiler for spatio-temporal mapping of dataflow graph (DFG) nodes, as illustrated in Figure~\ref{fig:bg_map}. 
This mapping includes assigning operations to CGRA PEs and routing data dependencies through configurable switches and registers. 
A common strategy, loop pipelining, allows concurrent scheduling of operations from different iterations, enhancing the kernel's throughput as shown in Figure~\ref{fig:bg_sch}. 
This process, also known as modulo scheduling, can be achieved by mapping DFG onto a modulo spatio-temporal resource graph, known as the Modulo Routing Resource Graph (MRRG) (Figure~\ref{fig:bg_mrrg})~\cite{rau1994iterative}.

In this study, we direct our attention to the optimization of ML workloads on user-defined CGRAs, employing the Morpher tool chain—a comprehensive design tool developed for CGRA modeling and compilation.
Morpher is an open-source framework that provides comprehensive support for modeling diverse CGRA architectures, supporting complex kernels, and verifying functionality. 
It enables users to design architecture characteristics through its architecture description language (ADL) and proficiently maps complex compute kernels. 
Morpher also auto-generates Verilog RTL for custom CGRAs, and validates functionality through Verilator-based simulations~\cite{verilator, pillars}. Hosted on GitHub, it integrates workflows for compilation, RTL generation, simulation, and verification, while incorporating Continuous Integration (CI) workflows to ensure error-free code and tested use cases.

The organization of this paper is as follows: Section~\ref{sec:framework_overview} provides an overview of Morpher framework. In Section~\ref{sec:CGRA_model}, we detail how the target CGRA is modeled. Section~\ref{sec:accelerating_ML} not only outlines our approach to accelerate ML kernels on CGRAs, but also presents evaluations of the applied optimizations and verification process of their execution.

\section{Morpher Framework Overview}
\label{sec:framework_overview}

\begin{figure}[t]
	\scriptsize
	\centering
    \includegraphics[width=\linewidth]{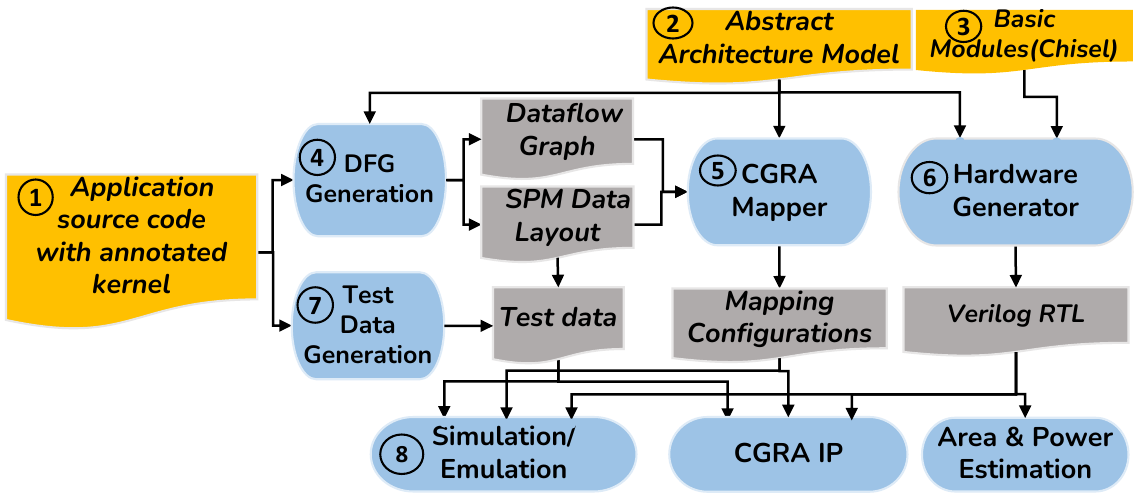}
	\caption{Overview of Morpher Framework}
	\label{fig:framework}
\end{figure}

Fig.~\ref{fig:framework} illustrates the overall Morpher framework. 
The pieces of the framework are numbered for easy reference.
Yellow pieces represent user-provided inputs, blue pieces represent the functional components, and grey ones represent intermediate results generated by the functional components.
The framework has three inputs: application source code with annotated kernel \circled{1}, the abstract architecture model \circled{2}, and a library of hardware description of basic CGRA modules \circled{3}.
The main components of the framework are Data-Flow Graph (DFG), and data layout generation \circled{4}, CGRA Mapper \circled{5}, hardware (RTL) generation \circled{6}, test data generation \circled{7}, simulation and emulation~\circled{8}.

CGRAs target loop kernels where the application spends a significant fraction of the execution time. 
The DFG generator \circled{4} is an LLVM-based pass that extracts the DFG of the target loop annotated in the application source code. 
Additionally, it constructs the multi-bank data layout by allocating the variables in the loop kernel to the on-chip memories of the target CGRA. 
The CGRA mapper \circled{5} maps the extracted DFG onto the CGRA fabric to maximize parallelism by exploiting intra- and inter-iteration parallelism with software pipelining (i.e., modulo scheduling)~\cite{rau1994iterative}. 
Morpher ADL supports a rich set of primitive constructs that model functional units,  register files, complex software-defined routers, and multi-banked memories accessible via shared bus interfaces. 
The mapper models the CGRA as a time-extended resource graph called MRRG~\cite{mei2002dresc} where the nodes of the DFG are mapped to the time-space resource instances to maximize throughput and minimize data routing cost. 
The resultant mapping configuration file describes the configuration for each resource cycle-by-cycle.

The architecture generator \circled{6} generates the Verilog RTL of the target CGRA design based on the user-provided abstract architecture model and the library of basic CGRA modules written in Chisel~\cite{pillars}. 
The test data generator \circled{7} for an application creates the data required for simulation and verification of the application execution. 
Finally, the simulator and emulator \circled{8} use the mapping configurations, the test data, and Verilog RTL to simulate and emulate the execution of the application on the specified architecture. 

\section{Modeling the target CGRA architecture}
\label{sec:CGRA_model}

For this study, the target CGRA is designed with an 8$\times$8 PE array, comprising 8 data memories  connected to the boundary PEs located on the left and right sides.
The CGRA is logically structured into four clusters, with each cluster accommodating a 4x4 PE array and two 8kB memory banks, in line with a 16-bit data path, as shown in Figure~\ref{fig:cgra}. This arrangement is described using the abstract ADL of Morpher, a flexible tool in .json format adept at capturing a range of CGRA architectures. 

Morpher's ADL balances high abstraction for user-friendliness with the need to handle intricate architectural specifics vital for verilog RTL generation. 
It does this by incorporating a library of crucial CGRA hardware modules like ALUs, LSUs, register files, multiplexers, and memory units, all developed in the Chisel language. 
This lets users tailor optimized architectures. 
Consequently, Morpher streamlines the design process, translating the ADL into an scala based Intermediate Representation (IR) that forms the Chisel top design and verilog RTL.

\section{Accelerating ML kernels on CGRA}
\label{sec:accelerating_ML}



In this section, we illustrate our strategy for accelerating diverse ML workloads, focusing primarily on the General Matrix Multiply (GEMM) and Convolution (CONV) kernels, using the Morpher toolchain. 
These kernels, despite being merely two examples among many ML kernels, act as crucial components in a plethora of ML models, contributing significantly to layers such as Fully Connected (FC), Convolution, Transformer models, LSTM, GRU, Bilinear, Self-Attention, and Graph Neural Networks (GNN) layers. 
While our methodology is demonstrated using GEMM and CONV kernels, it maintains broad applicability to numerous ML kernels on user defined CGRAs. 
Our attention is centered on diverse optimization strategies, including loop tiling, unrolling, and loop coalescing, which when combined, facilitate improved utilization of the CGRA resources and substantially boost performance.

\subsection{Tiling Strategy for ML Kernels}



The GEMM and CONV (single-input multiple-output channels) kernels are implemented on CGRA in a tiled manner using an output stationary dataflow, as shown in Listing~\ref{list:gemm_tiling} and ~\ref{list:conv_tiling}.
This is just one instances of many tiling techniques widely explored for spatial accelerators and effectively applicable to CGRAs~\cite{dave2019dmazerunner, moon2021evaluating}.

At the single CGRA level (lines 9-12 in Listing~\ref{list:gemm_tiling} and lines 9-16 in Listing~\ref{list:conv_tiling}), a specific-sized kernel, called ``TILE,''
 is mapped to an individual CGRA cluster. 
For GEMM, a matrix multiplication of size $TI \times TK \times TJ$ is mapped to a CGRA cluster. For CONV, a convolution of a tile of size $TO1 \times TO2 \times TCo$ with a filter of size $K \times K$ is carried out.
Each cluster computes an output matrix (O) with weights (W) and an input matrix (I) as inputs, the sizes of which are governed by the capacity of on-chip memory banks within each CGRA cluster. 
This mapping process is facilitated by the Morpher tool chain, detailed further in the next section.

The sequential loops manage data transfer from off-chip to on-chip memory, while computation is handled by parallel loops mapped onto CGRA clusters (lines 2-7 in Listing~\ref{list:gemm_tiling} and Listing~\ref{list:conv_tiling}). 
At the CGRA cluster level, data parallelism among different output tiles is leveraged. 
This allows multiple tiles to be spatially mapped on the CGRA cluster array, with each CGRA computing a single output tile. 
At the off-chip to on-chip level, any data exceeding the capacity of the on-chip memory banks is stored in off-chip memory, ensuring efficient data management throughout the system.




\begin{lstlisting}[language=c, caption=GEMM loop tiling and dataflow, label=list:gemm_tiling]
// Sequential loop: from off-chip to on-chip
for m in range(M/(TI*X)):
for n in range(N/(TJ*Y)):
for k in range(K/(TK)):
// Parallel loop: CGRA clusters
  for x in range(X):
  for y in range(Y):
// Single CGRA level
    for i in range(TI):
    for j in range(TJ):
    for k in range(TK)://map this
      O[][] += W[][]* I[][];
\end{lstlisting}

\begin{lstlisting}[language=c, caption=CONV loop tiling and dataflow, label=list:conv_tiling]
//Sequential loop: from off-chip to on-chip
for i.0 in range (O1/ X*TO1):
for j.0 in range (O2/ Y*TO2):
for c.0 in range(Co/ TCo):
// Parallel loop: CGRA clusters
  for x in range(X):
  for y in range(Y):
// Single CGRA level 
    for i in range(TO1):
    for j in range(TO2):
    for c in range(TCo):
        temp = 0;
        for k1 in range(K):
        for k2 in range(K):// map this:
            temp += I[] * W[];
    O[] = temp;
\end{lstlisting}



\subsection{Micro Kernel Mapping on CGRA}


\begin{table*}
\caption{Performance comparison of different kernels on target CGRA with speedup compared to base kernels}
\centering
\resizebox{\linewidth}{!}{    
\begin{threeparttable}    
\begin{tabular}{|l|r|r|r|r|r|r|r|}
\hline
\textbf{Kernel} & \textbf{Nodes} & \textbf{II (MII)} & \textbf{Utilization} & \textbf{\begin{tabular}[c]{@{}l@{}}Compute \\ time (ms)\tnote{*}\end{tabular}} & \textbf{\begin{tabular}[c]{@{}l@{}}Data transfer \\ time (ms)\tnote{*}\end{tabular}} & \textbf{\begin{tabular}[c]{@{}l@{}}Total execution \\ time (ms)\tnote{*}\end{tabular}} & \textbf{Speedup} \\ \hline
\textbf{GEMM} & 26 & 4 (4) & 40.63\% & 0.56 & 2.13 & 2.69 & 1$\times$ \\ \cline{1-8} 
\textbf{GEMM-U} & 58 & 6 (4) & 60.42\% & 0.25 & 2.13 & 2.38 & 1.1$\times$ \\ \cline{1-8} 
\textbf{GEMM-U-C} & 79 & 8 (8) & 61.72\% & 0.27 & 0.49 & 0.76 & 3.5$\times$ \\ \cline{1-8} 
\textbf{CONV} & 27 & 4 (4) & 42.19\% & 8.32 & 306.38 & 314.70 & 1$\times$ \\ \cline{1-8} 
\textbf{CONV-U-C-1} & 100 & 12 (7) & 52.08\% & 1.53 & 12.75 & 14.28 & 22$\times$ \\ \cline{1-8} 
\textbf{CONV-U-C-2} & 153 & 11 (10) & 86.93\% & 1.26 & 11.19 & 12.45 & 25.2$\times$ \\ \hline
\end{tabular}
\begin{tablenotes}
\item [*] The evaluation is conducted at a 100 MHz CGRA frequency and a 50 MBps host-to-CGRA data transfer rate. 
The primary focus of this study is to evaluate performance enhancement through compilation, not striving for the maximum performance achievable through efficient RTL silicon implementation, which is currently in the development phase. 
\end{tablenotes}
\end{threeparttable}
}
\label{tab:perf_comp}
\end{table*}

In this section, we explore the process of optimizing and mapping GEMM and CONV kernels, onto a single CGRA cluster. 
We further analyze the performance implications of these optimizations.
The evaluated GEMM and CONV kernels have dimensions of $64^3$ and $64^3 \times 3^2$, respectively. Their corresponding tile sizes fitting into onchip memory banks of single CGRA cluster are $64 \times 16 \times 64$ for GEMM and $64^2 \times 1 \times 3^2$ for CONV.

Both GEMM and CONV kernels consists of nested loops.
The user only needs to provide the application C source code to the Morpher toolchain and annotate the innermost loop that should be mapped onto the CGRA, here annotated as "map this" (Line 11 in Listing \ref{list:gemm_tiling} and line 14 in Listing~\ref{list:conv_tiling}).
The toolchain then generates a dataflow graph, representing the innermost loop body, and maps it onto the CGRA cluster. 
This mapping generates the necessary configurations to exploit the parallel computational capacity of the CGRA for executing the kernel. 
The toolchain also manages data layout in memory banks, mapping data arrays onto them to synchronize computation and data mapping.

Table \ref{tab:perf_comp} provides a performance evaluation. The Initiation Interval (II) represents the cycle count between start of two consecutive iterations, while the Minimum II (MII) is the smallest possible II dictated by the CGRA resource and loop's recurrence constraints~\cite{recaware}. 
For both the base GEMM kernel (26 DFG nodes) and the base CONV kernel (27 DFG nodes), total execution times are 2.69~ms and 314.70~ms. 
In both cases, Morpher succeeds in achieving the theoretical MII of 4. 
Notably, the lower performance of the CONV kernel arises due to an increased kernel invocation overhead, which includes transferring outer loop iteration variables from the host processor to the CGRA, as well as extended pipeline draining time. The latter refers to the period during which the pipeline completes executing instructions after the final loop iteration has commenced. These factors are amplified due to the CONV kernel's higher number of nested loop levels (5 compared to GEMM's 3).
This overhead, combined with less than optimal resource utilization (40\% for GEMM and 42.19\% for CONV), spotlights the opportunities in optimizing CGRA performance for complex ML kernels.

Incorporating loop unrolling optimization into the GEMM kernel, as shown in Listing \ref{list:gemm2}, significantly elevates performance. 
This optimization inflates the number of operations within the loop body, hence increasing the number of DFG nodes and amplifying parallelism, which optimizes the utilization of the CGRA PEs. As a result, the unrolled version (GEMM-U) demonstrates an increased DFG nodes count from 26 to 58 and an enhancement in utilization from 40\% to 60\%. This culminates in a decrease in computation time from 0.56ms to 0.25ms. 
These reductions confirm that loop unrolling efficiently enhances the compute utilization, resulting in an  an improved performance, which is 1.13$\times$ better than the base kernel.

\newpage

\begin{lstlisting}[language=C, caption=Unrolled GEMM kernel (GEMM-U), label=list:gemm2]
for (i=0;i<TI; i++)
for (j=0;j<TJ; j++)
for (k=0;k<TK; k=k+4): //map this
  O[i][j] += W[i][k]* I[k][j]+ W[i][k+1]* I[k+1][j]
    W[i][k+2]* I[k+2][j]+W[i][k+3]* I[k+3][j];
\end{lstlisting}

Loop coalescing significantly enhances the efficiency of the CGRA implementation, by reducing invocation overheads and pipeline draining time. This is clearly demonstrated in the results for the GEMM-U-C (Listing~\ref{list:gemm_u_c}) and the CONV-U-C (Listing~\ref{list:conv_u_c})  kernels.
The GEMM-U-C kernel coalesces all three loops, resulting in a DFG with 79 nodes and an II of 8. 
This kernel requires only a single loop invocation per CGRA cluster to complete $64^3$ kernel size.
The data transfer time is reduced to 0.49~ms, culminating in a total execution time of 0.76~ms. This significantly enhances the overall performance, as evidenced by a performance boost of 3.54 times compared to base kernel.
Similarly, the CONV kernel also shows marked improvement when optimized. 
The CONV-U-C-1 kernel, which coalesces the innermost two loops and fully unrolls them when K=3, results in a DFG with 100 nodes and an II of 12. 
The compute time is significantly reduced to 1.53~ms, as is the data transfer time to 12.75~ms, yielding a total execution time of 14.28~ms. 
This optimization leads to an impressive performance increase to 22.03$\times$ compared to the base kernel.

Finally, the CONV-U-C-2 kernel, which coalesces all five loop levels, demonstrates a further improvement. 
This kernel necessitates 16 invocations per CGRA cluster to complete $64^3 \times 3^2$ kernel size.
It results in a DFG with 153 nodes and an II of 11 with 86\% utilization. 
This optimization results in a performance boost of 25.28$\times$ compared to the base CONV implementation. 
These findings underscore the vital role and efficacy of loop coalescing in achieving significant performance gains in CGRA implementations.

\begin{lstlisting}[language=C, caption=Unrolled \& coalesced GEMM kernel (GEMM-U-C), label=list:gemm_u_c]
for (n=0;i=0;j=0;k=0;n<TI*TJ*TK; n++){: //map this
  O[i][j] += W[i][k]*I[k][j]+W[i][k+1]*I[k+1][j]
  W[i][k+2]*I[k+2][j]+W[i][k+3]*I[k+3][j];k = k + 4;  
  if(k+1 >= TK) {k=0; ++j;}
  if(j == TJ) {j=0; ++i;}}
\end{lstlisting}

\begin{lstlisting}[language=C, caption=Unrolled \& coalesced CONV kernel (CONV-U-C-2), label=list:conv_u_c]
for (int ijk=0;ijk<TCo*TO1*TO2; ijk++){: //map this
  O[] = I[] * W[] + I[] * W[] + I[] * W[]
  + I[] * W[] + I[] * W[] + I[] * W[]
  + I[] * W[] + I[] * W[] + I[] * W[]; j = j + 1;
  if(j+1 > O2){j=0;++i;}
  if(i == O1){i=0;++c;}}
\end{lstlisting}

\subsection{Functional Verification}



Morpher simplifies the task of generating application test data for the simulation of loop kernels, an indispensable component of CGRA functional verification. It instruments the application source code by inserting data recording functions to capture the live-in (I, W, O arrays, iteration variables transferred from outermost loops) and live-out (output O array) variables of the target loop kernel. This instrumented program is then run on a general-purpose processor, and the variables are recorded as test data for use by the simulator. In the ensuing simulation and verification phase, the Chisel top design of the target CGRA is simulated using Verilator and Chisel I/O testers, with the CGRA model functioning as a memory-mapped slave device to a host processor. The live-in variables from the recorded test data are loaded into each memory unit, and the mapping configurations from the mapper are uploaded into the automatically generated control modules. The simulator then carries out the operations, routing data through multiplexers, operating on the functional units, and recording the results to registers and memories, all as per the mapping configurations. 
The post-simulation memory content is finally compared with the expected results, validating the CGRA functionality with Morpher generated configurations.

\section{Conclusion and future works}

CGRAs backed by the efficient kernel mapping of the Morpher toolchain, offer a promising route for ML application acceleration. 
In our future work, we aim to merge the Morpher toolchain with MLIR's high-level compilation front-end. 
This integration will automate optimization techniques, further exploring the CGRA design space, and enhancing performance. 
This effort continues to strive towards unlocking the full potential of CGRA technology.

\section{Acknowledgment}
This work was partially supported by the National Research Foundation, Singapore under its Competitive Research Programme Award NRF-CRP23-2019-0003 and Singapore Ministry of Education Academic Research Fund T1 251RES1905.

\end{document}